\documentclass[lettersize,journal]{IEEEtran}
\usepackage{amsmath,amsfonts}
\usepackage{array}
\usepackage{textcomp}
\usepackage{url}
\usepackage{graphicx}
\usepackage{cite}
\usepackage{booktabs}
\usepackage{balance}
\newcommand{\kdse}{KD-DSE}
\graphicspath{{./figure/}}
\begin{document}

\title{\kdse{}: A Knowledge-Driven LLM Framework for Fast Microarchitecture Design Space Exploration}

\author{Siqing~Fu, Xin~Liu, Yuhan~Tang, Hanqing~Li, and Feng~Guo
\thanks{Manuscript received XXXX; revised XXXX.}
\thanks{Siqing Fu and Feng Guo are with the Jiangnan Institute of Computing Technology, Wuxi, Jiangsu, China (e-mail: fusiqingnudt@nudt.edu.cn; gf90966@163.com). (Corresponding author: Feng Guo.)}
\thanks{Xin Liu is with the National Research Center of Parallel Computer Engineering and Technology, Beijing, China (e-mail: yyylx@263.net).}
\thanks{Yuhan Tang and Hanqing Li are with the National University of Defense Technology, Changsha, Hunan, China (e-mail: tangyuhan@nudt.edu.cn; lihanqing23013@nudt.edu.cn).}}

%\markboth{IEEE Computer Architecture Letters}%
%{Fu \MakeLowercase{\textit{et al.}}: \kdse{}: A Knowledge-Driven LLM Framework for Fast Microarchitecture Design Space Exploration}

\maketitle

\begin{abstract}
Domain-specific architectures have made design space exploration (DSE) a bottleneck of processor design, yet each candidate costs a full simulation, so exhaustive search is out of reach. Surrogate models need thousands of simulations before they predict reliably, and LLM agents that search directly hallucinate, lack cold-start architectural knowledge, and follow a fixed per-round policy. We present \kdse{}, a knowledge-driven LLM framework for fast microarchitecture DSE. \kdse{} couples a five-state constrained agent with real gem5-McPAT evaluation, retrieves architect-curated guidance from an Architecture Knowledge Base conditioned on the current search state and bottleneck diagnosis, and allocates its budget with a diagnosis-guided two-phase exploration strategy. On a 29-parameter out-of-order core with a budget of 18 simulations, \kdse{} achieves about 4.8\% higher hypervolume and about 21.7\% lower ADRS than the state-of-the-art MicroEvo, and the ablation attributes the gain chiefly to the Knowledge Base. These results indicate that constraining LLM reasoning and grounding it in architectural knowledge can drive fast and effective microarchitecture DSE. The code will be open-sourced upon acceptance.
\end{abstract}


\section{Introduction}
\label{sec:intro}

\IEEEPARstart{W}{ith} the end of Dennard scaling and the slowing of Moore's law, application-specific customization through domain-specific architectures (DSAs) has become the default path to continued performance growth \cite{alphachip2021,lpcm2026}. Each emerging workload, from large-model inference to edge devices, demands a new design point in a microarchitecture space that spans pipeline widths, instruction windows, cache hierarchy, and memory controller; a space of 29 parameters already holds \(1.5\times10^{17}\) configurations. Because every candidate costs a full simulation, design space exploration (DSE) has long been the bottleneck of the processor design cycle.

Prior work accelerates DSE with heuristic search \cite{nsgaii2002}, Bayesian optimization \cite{tpe2011}, reinforcement learning \cite{archgym2023}, and learned surrogate models \cite{boomexplorer2021,trendse2023}.
% floats declared during page-1 assembly: framework lands in a page-2 column, statemachine spans the page-2 top
\begin{figure}[!t]
\centering
\includegraphics[width=0.92\columnwidth]{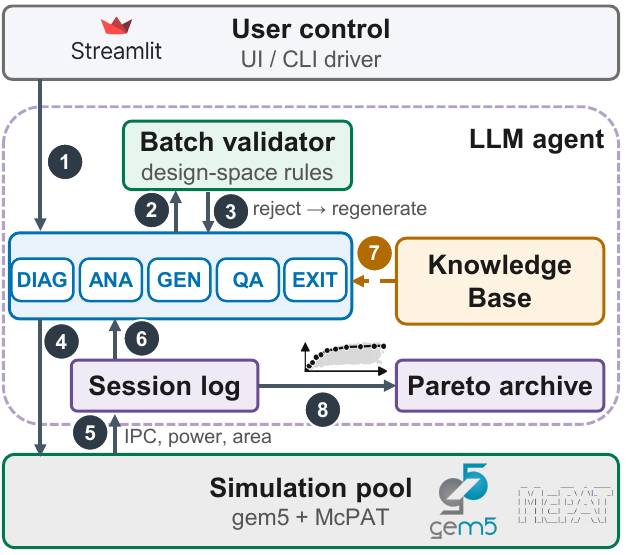}
\caption{The \kdse{} framework.}
\label{fig:framework}
\end{figure}
\begin{figure*}[!t]
\centering
\includegraphics[width=0.98\textwidth]{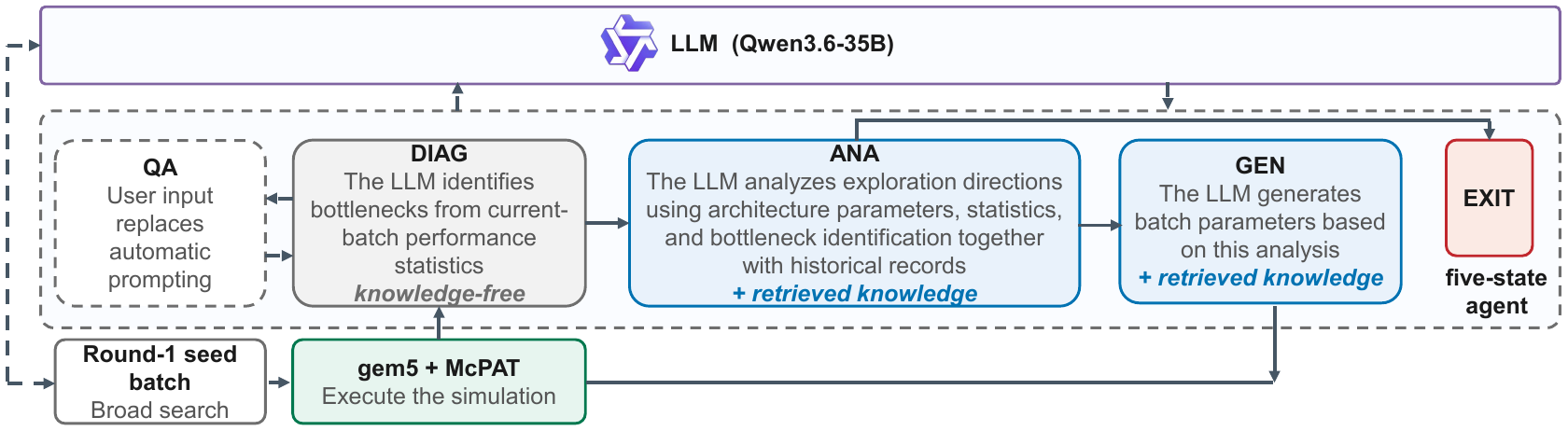}
\caption{The five-state agent of \kdse{}.}
\label{fig:statemachine}
\end{figure*} All of these treat the simulator as a black box and consume thousands of labeled simulations before they predict reliably---precisely the budget industrial practice cannot spare---and in the small-sample regime their predictions drift from ground truth and misdirect the search \cite{microevo2026}. This regime instead needs transferable architectural priors and reasoning, which large language models (LLMs) can supply, and recent systems therefore embed LLMs directly in the DSE loop \cite{gem5copilot2025,LEMOE2025,chata2_2026,microevo2026,agentdse2026}. Yet three problems remain unresolved. First, LLM reasoning is prone to hallucination: free-running agents drift across tens of turns and produce outputs that violate the design space or the schema, and systems that avoid this keep an engineer in the loop \cite{gem5copilot2025}. Second, general-purpose LLMs are not trained on architectural parameters and their performance--power--area results, so at cold start they cannot tell which regions of the space are worth probing; the closest exception, MicroEvo \cite{microevo2026}, injects knowledge only during the search, so what it knows grows with the iteration count. Third, DSE needs a broad sweep first and a focused search later, but existing agents control every round with a fixed strategy and no structured diagnosis, wasting budget where the marginal gain is small and finding the knee only by accident.

To address these problems, we present \kdse{}, a knowledge-driven LLM framework for fast processor DSE. A five-state agent proposes constraint-compliant design batches and evaluates them on real gem5-McPAT simulations. A state-driven Architecture Knowledge Base (KB), grounded in simulation experience and architect expertise, conditions on the current search state and bottleneck diagnosis to deliver targeted, benchmark-agnostic guidance. A diagnosis-guided two-phase strategy anchors both extremes of the cost range with a seed sweep, then composes each batch from the focus zone and the diagnosis topics, concentrating the budget where the performance-per-cost slope is steepest. On mmbench, \kdse{} outperforms the state-of-the-art MicroEvo \cite{microevo2026} with both higher hypervolume (HV, $0.6843$ vs.\ $0.6529$) and lower average distance to the reference set (ADRS, $0.1011$ vs.\ $0.1291$), and the advantage holds across benchmarks. The contributions of this letter are:
\begin{enumerate}
\item a five-state agent that drives concurrent gem5-McPAT evaluations through a diagnosis, analysis, and generation loop, keeping every candidate within the design space and schema;
\item a Knowledge Base that retrieves architect-curated entries by the current cost region and diagnosis, providing reliable guidance from the first round without online accumulation;
\item a two-phase exploration strategy that anchors the cost range by a forced seed sweep and then refines toward the knee using focus-zone and diagnosis signals.
\end{enumerate}

\section{\kdse{} Design}
\label{sec:design}

% fig_compare declared during page-2 assembly so it floats to the top of page 3
\begin{figure*}[!t]
\centering
\includegraphics[width=0.96\textwidth]{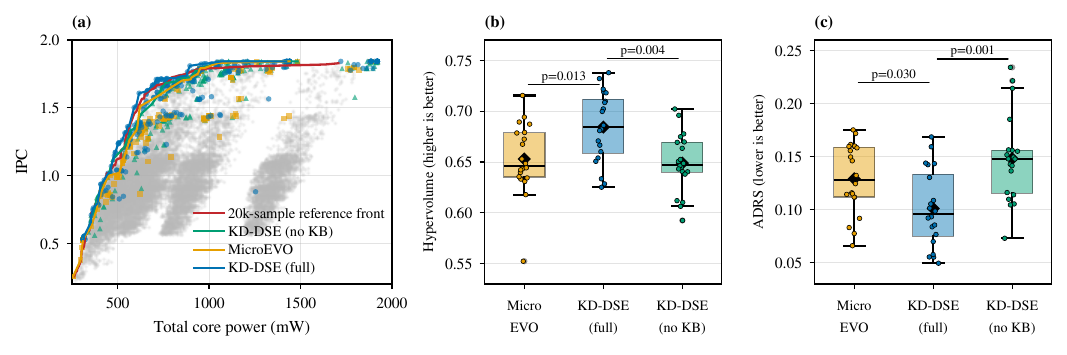}
\caption{Comparison over 20 runs per configuration. (a) All measured points against the $20{,}658$-point random-sampling library; lines show each group's pooled non-dominated front, and the red staircase is the library reference front. (b) Per-run HV and (c) ADRS distributions (diamonds: means; $p$: Holm-corrected permutation tests).}
\label{fig:compare}
\end{figure*}

\subsection{Framework Overview}
\label{sec:overview}
Fig.~\ref{fig:framework} shows the \kdse{} framework, which addresses the three problems with three components: the five-state agent constrains the process, the state-driven Knowledge Base (KB) supplies knowledge, and the two-phase strategy spends the budget. The user configures the mission, the design space, and the budget (1). The agent emits batches of at most $n_{\mathrm{sim}}$ designs (2); a validator returns illegal batches for regeneration (3); legal batches run on concurrent gem5-McPAT chains (4); the measured IPC, power, and area enter the session log (5) and supply diagnosis evidence and history at the next turn (6); and the KB retrieves a block into the reasoning and generation turns (7). On EXIT, the archive and the HV/ADRS report are produced (8). A static system prompt fixes the role, global rules, state semantics, and mission strategy, while a per-turn user prompt carries the current state, goal, and log, so each turn receives the evidence and instructions it needs.

\subsection{Five-State Constrained Agent}
\label{sec:agent}
Fig.~\ref{fig:statemachine} illustrates the five-state agent. At each turn the agent returns one schema-constrained response for exactly one of five states under a fixed protocol. \emph{DIAG} diagnoses the bottleneck: it names the module and parameter behind the loss, citing measured evidence, and may not propose configurations. \emph{ANA} builds the strategy: it checks whether the previous prediction was confirmed or refuted, reasons step by step, and chooses to consolidate, explore, or balance. \emph{GEN} emits the next batch of at most $n_{\mathrm{sim}}$ designs, the only output that can trigger simulations; \emph{QA} answers user questions without a batch; \emph{EXIT} submits the final design. The turn order is fixed as DIAG\,$\rightarrow$\,ANA\,$\rightarrow$\,GEN, which keeps the stages of design-parameter reasoning apart.

Limiting the states alone does not stop hallucination, so three checks guard the loop. First, illegal parameters: batches that violate the design-space rules are sent back to ANA with an error message for regeneration. Second, unjustified termination: a hard exit is allowed only when the exit conditions hold, such as three consecutive rounds without a new frontier, or when the budget is exhausted. Third, repeated designs: the agent compares each batch with the session log before simulation and drops duplicates, so no simulation slot is wasted.

\subsection{Architecture Knowledge Base and State-Driven Retrieval}
\label{sec:knowledge}
The Architecture Knowledge Base is the only place that stores external architectural knowledge in \kdse{}. Following the retrieval-augmented generation paradigm \cite{rag2020}, it performs deterministic, state-driven selection over curated rules. The rules are distilled from a large body of prior simulation runs and the experience of human architects, and each records how a structural choice relates to design quality. A rule is value-independent: it compares a parameter only against its own range, stating the direction of change and which candidate is better at equal cost, so the same rule transfers across modules, sizes, and workloads. For instance, at equal cost the smallest cache is optimal, so increasing cache capacity pays only in the high-performance, high-cost range; and since Pareto designs tend to be asymmetric, the generator raises one pipeline part at a time while keeping the all-minimum design, itself a frontier point, in the seed batch.

Retrieval is a deterministic tag match over entry metadata, driven by two state inputs: the focus zone of the measured frontier and the bottleneck confirmed by the latest diagnosis. The retrieved block is built in three layers. The \emph{CORE} layer holds the foundational rules and is injected on every turn, so they cannot be rotated out as the focus moves. The \emph{ZONE} layer adds entries tagged with the focus zone (\emph{cheap}, \emph{mid}, or \emph{high}), taken as the cheapest zone that is still under-sampled or still holds a large unmeasured gap. The \emph{TOPIC} layer adds entries whose tags intersect those of the latest diagnosis, so a diagnosis attributing the problem to the memory hierarchy retrieves memory knowledge. The layers are combined and truncated at a fixed word budget, which bounds the prompt length and avoids the hallucination that overly long prompts induce.

\subsection{Two-Phase Exploration Strategy}
\label{sec:strategy}
Existing LLM-driven DSE frameworks search the same way in every round, but DSE requires a wide scan first and a narrower search later. \kdse{} therefore splits the search into two phases. In the first phase, a one-time seed sweep generates the designs with the best performance and the lowest cost through prompt constraints, then distributes the remaining budget of the batch across several cost levels; the purpose is to cover the front broadly and to establish the cost range that guides the refined search. In the second phase, the round-by-round search refines upon the sweep: DIAG distinguishes which cost ranges remain under-sampled, diagnoses the current bottleneck within that range, and supplements the search for that range. The sweep fixes the cost range in the first round, and the diagnosis-driven rounds that follow push the sampled points as close to the Pareto front---and its knee, where cost-effectiveness meets high performance---as the budget allows.

\section{Evaluation}
\label{sec:exp}

\subsection{Experimental Setup}
\label{sec:setup}
The experiments run on gem5 v25.1.0.1 \cite{gem52011} and McPAT v1.3 \cite{mcpat2009}, where gem5 statistics feed the McPAT templates that produce power and area. The agent is powered by Qwen3.6-35B \cite{qwen3_2025} served locally by vLLM \cite{vllm2023} to ease reproduction. We use the O3 core of gem5 over a full-core space of 29 parameters spanning the core, the L1I and L1D caches, the L2 cache, and the memory controller; rules that remove infeasible combinations prune the raw $1.5\times10^{17}$ configurations to $7.1\times10^{16}$. All evaluations run on a blocked mmbench \cite{gem5copilot2025} with a $128\times128$ matrix and a block size of 32. Because sweeping the space is prohibitive, $20{,}658$ random configurations form the evaluation baseline, whose objective front $P^{*}$ has 128 points. We measure front quality with hypervolume (HV), the volume dominated by the evaluated set up to a reference point in normalized log-objective space, where higher is better, and with ADRS, the average over $P^{*}$ of the worst-objective relative excess of the closest evaluated point, where lower is better.

Each run uses a budget of 18 configurations---three rounds of six concurrent gem5-McPAT chains---chosen to measure how quickly a framework approaches the front under expensive simulation; we execute 20 independent runs per configuration so that the stochasticity of the LLM averages out. We compare against the state-of-the-art MicroEvo \cite{microevo2026}, used unmodified on our space, workload, and stack, and against NSGA-II \cite{nsgaii2002} as the classical heuristic baseline; MicroEvo has been shown to outperform prior methods, including NSGA-II, BOOM-Explorer \cite{boomexplorer2021}, and LEMOE \cite{LEMOE2025}, under a comparable budget.

\subsection{Round-by-Round Behavior}
\label{sec:case}
A single \kdse{} search already shows the two-phase strategy at work. The first-round seed sweep produces the two corner designs---the highest-performance and the most cost-effective---and spans 311 to 1920\,mW of core power, a rough scan of the whole cost range; the seven pipeline widths share one symmetric structure in which all widths grow together. From these results the agent learns that the 670--964\,mW slope is steep and needs finer search. From the second round on, the search region narrows: since near-front configurations are often asymmetric, the second round raises one width at a time and reduces the L2 size toward the lowest level, which clearly improves IPC at the same power, lifting HV from $0.6562$ to $0.7274$ and cutting ADRS from $0.1500$ to $0.0584$. The configuration at 1.709 IPC and 711\,mW is the final knee, beyond the random-search front. The third round keeps the same strategy and densifies the search near the knee; every point moves closer to the front, although no new knee appears. Of the final 12 front points, 2 come from the first round, 5 from the second, and 5 from the third.

\begin{table}[!t]
\caption{Main Results Over 20 Independent Runs (Mean$\pm$Std, Best in Parentheses)}
\label{tab:main}
\centering
\footnotesize
\setlength{\tabcolsep}{4pt}
\begin{tabular*}{\columnwidth}{@{\extracolsep{\fill}}lcc@{}}
\toprule
Configuration & HV $\uparrow$ & ADRS $\downarrow$ \\
\midrule
MicroEvo \cite{microevo2026} & 0.6529$\pm$0.0351 (0.7154) & 0.1291$\pm$0.0329 (0.0660) \\
NSGA-II \cite{nsgaii2002} & 0.6032$\pm$0.0411 (0.6696) & 0.1853$\pm$0.0450 (0.1024) \\
\kdse{} (no KB) & 0.6489$\pm$0.0289 (0.7021) & 0.1482$\pm$0.0410 (0.0730) \\
\kdse{} (full) & \textbf{0.6843$\pm$0.0348} (\textbf{0.7378}) & \textbf{0.1011$\pm$0.0364} (\textbf{0.0495}) \\
\bottomrule
\end{tabular*}
\vspace{2pt}
\par\noindent\footnotesize vs.\ MicroEvo: $\Delta$HV $+4.8\%$ ($p{=}0.013$, $d{=}0.90$), $\Delta$ADRS $-21.7\%$ ($p{=}0.030$, $d{=}0.81$); vs.\ NSGA-II: $\Delta$HV $+13.4\%$ ($p{<}0.001$, $d{=}2.16$), $\Delta$ADRS $-45.4\%$ ($p{<}0.001$, $d{=}2.06$); vs.\ no-KB: $\Delta$HV $+5.5\%$ ($p{=}0.004$, $d{=}1.11$), $\Delta$ADRS $-31.8\%$ ($p{=}0.001$, $d{=}1.21$). Holm-corrected permutation tests, Cohen's $d$.
\label{tab:stats}
\end{table}

\subsection{Comparison with the State of the Art}
\label{sec:comparison}
Fig.~\ref{fig:compare}(a) summarizes the results: \kdse{} produces the best front overall, outperforming both MicroEvo and NSGA-II. MicroEvo builds architectural knowledge only as the search proceeds, so it starts each run without usable prior knowledge, whereas \kdse{} provides knowledge at the start and updates it through state-driven retrieval, so it can test promising regions from the first round. Figs.~\ref{fig:compare}(b) and \ref{fig:compare}(c) and Table~\ref{tab:main} quantify the gap: the mean HV of \kdse{} is 0.6843 against 0.6529 for MicroEvo, about 4.8\% higher, and its mean ADRS is 0.1011 against 0.1291, about 21.7\% lower. Raw $p$-values come from a permutation test corrected with the Holm--Bonferroni method; the corrected values, 0.013 for HV and 0.030 for ADRS, are both below 0.05, and Cohen's $d$ of 0.90 and 0.81 indicates large effect sizes, so \kdse{} outperforms MicroEvo significantly on both metrics. The margin over NSGA-II is larger still.

\begin{figure}[!t]
\centering
\includegraphics[width=0.96\columnwidth]{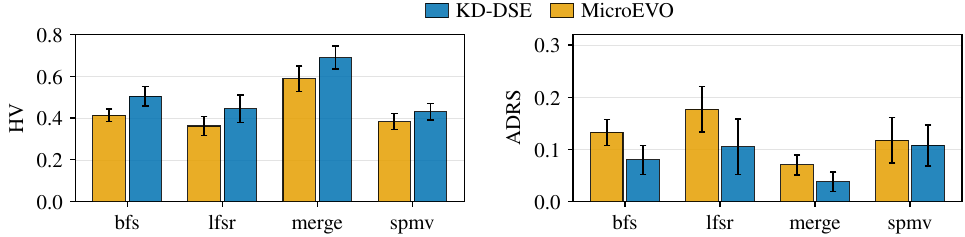}
\caption{HV and ADRS of \kdse{} versus MicroEvo \cite{microevo2026} across four microbenchmarks.}
\label{fig:morebench}
\end{figure}

\subsection{Cross-Benchmark Generalization and Ablation}
\label{sec:ablation}
We further evaluate \kdse{} on four microbenchmarks \cite{polyarch_cs251a_2018}---bfs, lfsr, merge, and spmv---which differ in memory locality, arithmetic intensity, and regularity. As no random sampling is available, the reference set for HV is the union of the fronts found by the two methods, averaged over twenty runs with the same budget. Fig.~\ref{fig:morebench} shows that \kdse{} achieves higher HV and lower ADRS on all of them, demonstrating generalization across benchmarks.

To isolate the source of the gain, we ablate the Knowledge Base under the same budget, since the five-state agent and the two-phase strategy keep the search valid and on budget regardless. The no-KB front in Fig.~\ref{fig:compare}(a) is clearly lower than the random-sampling reference and visually close to the MicroEvo result. Removing the KB drops the mean HV from 0.6843 to 0.6489 and raises the mean ADRS from 0.1011 to 0.1482, with corrected $p$-values of 0.004 and 0.001 and Cohen's $d$ of 1.11 and 1.21 (Table~\ref{tab:main}). The Knowledge Base is thus the main driver of \kdse{}'s search quality.

\section{Conclusion}
\label{sec:conclusion}
This letter presented \kdse{}, a knowledge-driven LLM framework for fast processor DSE. The five-state agent confines generation to machine-checked stages, the state-driven Knowledge Base supplies architect-curated, benchmark-agnostic guidance from the first round, and the two-phase strategy anchors the cost range and refines toward the knee. \kdse{} improves HV by 4.8\% and reduces ADRS by 21.7\% over the state-of-the-art MicroEvo and outperforms NSGA-II by a clear margin on both metrics. These results suggest that grounding and constraining LLM reasoning in architectural knowledge can drive rapid microarchitecture DSE.

\balance
\bibliographystyle{IEEEtran}
\bibliography{reb}

\end{document}